\documentclass[pre,aps,superscriptaddress,showpacs]{revtex4-1}
\usepackage{natbib}
\usepackage{amsmath}
\usepackage{epsfig}
 	
\begin{document}	
\title{Calculating work in weakly driven quantum master equations: backward and forward equations}
\author{Fei Liu}
\email[Email address: ]{feiliu@buaa.edu.cn}
\affiliation{School of Physics and Nuclear Energy Engineering, Beihang University, Beijing 100191, China}
\date{\today}

\begin{abstract}
{I present a technical report indicating that the two methods used for calculating characteristic functions for the work distribution in weakly driven quantum master equations are equivalent. One involves applying the notion of quantum jump trajectory [Phys. Rev. E 89, 042122 (2014)], while the other is based on two energy measurements on the combined system and reservoir [Silaev, et al., Phys. Rev. E 90, 022103 (2014)]. These represent backward and forward methods, respectively, which adopt a very similar approach to that of the Kolmogorov backward and forward equations used in classical stochastic theory. The microscopic basis for the former method is also clarified. {In addition, a previously unnoticed equality related to the heat is also revealed.}}
\end{abstract}
\maketitle

{\noindent \it Introduction.}
Recently, there has been growing interest in the heat and work of nonequilibrium quantum processes~\cite{Bochkov1977,Kurchan2000,Tasaki2000,
Yukawa2000,Mukamel2003,DeRoeck2004,Allahverdyan2005,
Talkner2007,DeRoeck2007,Andrieux2008,Talkner2009,Esposito2009,
Campisi2011,Subasi2012,Horowitz2013,DeRoeck2004,
Esposito2006,Crooks2008,Horowitz2012,Liu2012,Chetrite2012,
Hekking2013,Horowitz2013,Leggio2013,Liu2014,Liu2014a,Silaev2014,
Suomela2014,Suomela2015,Batalhao2014,ShuomingAn2015,Jarzynski2015}. Studies focusing on this issue have mainly been motivated by an interest in extending the important classical fluctuation relations into the quantum regime, e.g., the celebrated Bochkov-Kuzovlev equality~\cite{Bochkov1977} and the Jarzynski equality (JE)\cite{Jarzynski1997}.

\begin{figure}
\includegraphics[width=1\columnwidth]{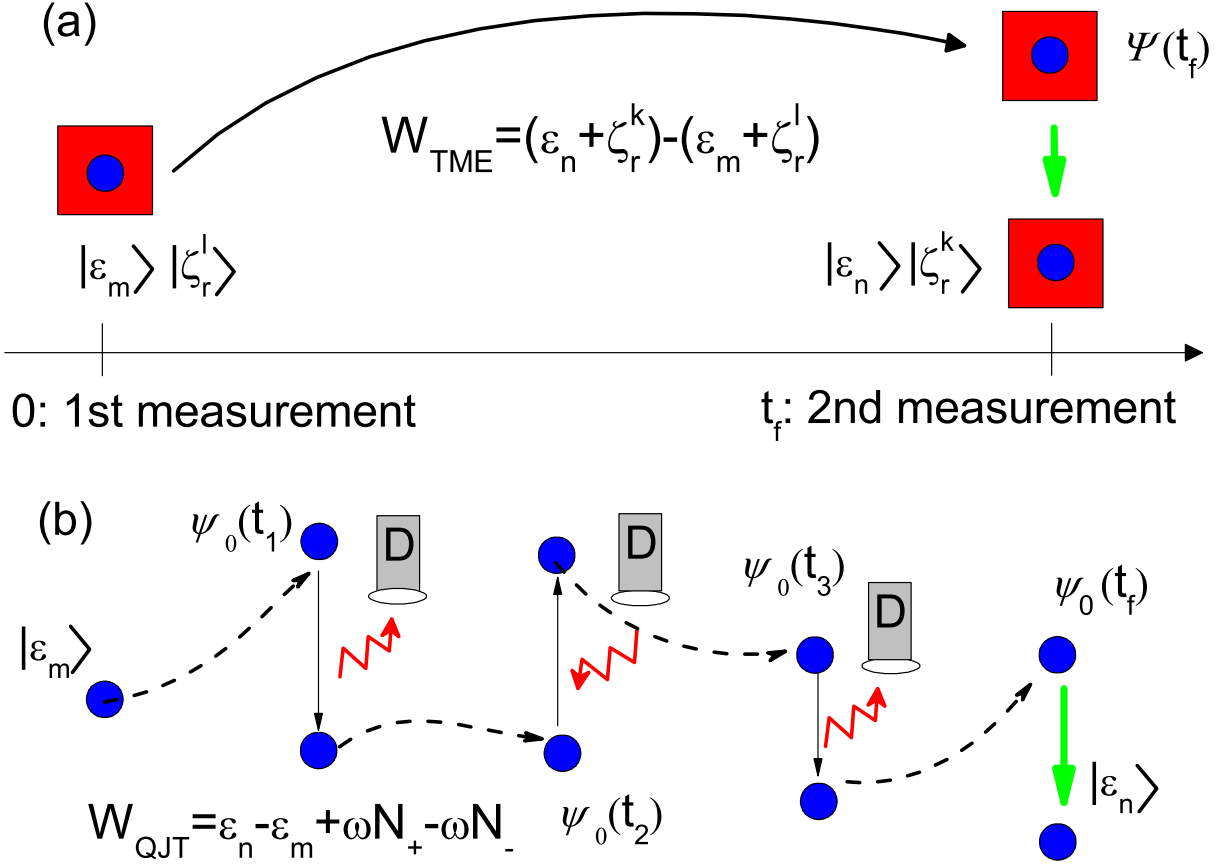}
\caption{{A schematic diagram describing the two definitions of work based on the TEM scheme on (a) the combined two-level quantum system (the blue circles) and reservoir (red squares)  and on (b) the QJT. In (a), the Hamiltonian, $H_0$, of a bare system has a discrete eigenvector and eigenvalue: $H_0|\varepsilon_n\rangle$$=$$\varepsilon_n|\varepsilon_n\rangle$. The Hamiltonian $H_r$ of the reservoir has the following eigenvector and eigenvalues: $H_r|\zeta_r^k\rangle$$=$$\zeta_r^k|\zeta_r^k\rangle$. The evolution of the wave vector, $\Psi(t)$, of the composite system is unitary under the whole Hamiltonian. The green arrow on the right-hand side denotes the projected energy measurements on the system and reservoir at time, $t_f$. In (b), the dashed lines denote the Schr\"{o}dinger-like evolution of the state vector, $\psi_0(t)$, of the system under a non-Hermitian Hamiltonian, while thin lines represent occasional jumps of the state vector due to emitting or absorbing a quantum $\omega$ of energy into or from the reservoir. These energies are recorded by the detectors labeled by the letter D~\cite{Breuer2002}. I assume their   total numbers to be $N_+$ and $N_-$, respectively. Notice that here the projection at time $t_f$ is conducted solely on the bare system; see the green line. }}
\label{figure}
\end{figure}

{Compared with their classical counterparts~\cite{Jarzynski2007}, which are physically intuitive, the definitions of fluctuating heat and work become very delicate in the quantum case. In order to formulate a quantum JE, in a closed quantum system, a two-energy measurements (TEM) scheme was proposed by Kurchan~\cite{Kurchan2000} to define the work. Although this  definition still faces criticisms related to the fact that the scheme may destroy the initial quantum-coherent superposition of the system~\cite{Allahverdyan2014}, it has been widely accepted in the field~\cite{Campisi2011,Esposito2009,Liu2014b}. Because closed quantum systems are not common in reality, there have also been many attempts to generalize this definition to include open quantum systems~\cite{Breuer2002}. These efforts can be roughly divided into two types of method. The first type~\cite{Talkner2009,Esposito2009,Silaev2014} involves the combination of the system and its surrounding heat reservoir as a composite system. The TEM scheme is then conducted on the system and reservoir. As the interaction between these is weak, the energy eigenvalue change obtained using the TEM for the system is referred to as the internal energy change, while the energy eigenvalue change obtained by the TEM for the reservoir is referred to as the heat released from the system, $Q_{TEM}$. Therefore, the work done on the open system, $W_{TEM}$, is the sum of the internal energy change and the heat. The second type of method is based on the quantum jump trajectory (QJT) that is unraveled from the Lindblad quantum master equations~\cite{DeRoeck2004,DeRoeck2007,Derezinski2008,Crooks2008,Hekking2013,Horowitz2012,Leggio2013,Liu2014,Liu2014a}. Under this notion, the energy change of the heat reservoir is continuously measured. This is interpreted as the released heat along a trajectory, $Q_{QJT}$. If one preserves the internal energy change of the system mentioned above, an alternative work, $W_{QJT}$,  is then the sum of the heat and internal energy change along the same trajectory. Figure~(\ref{figure}) schematically illustrates the difference between these two types of work in a two-level quantum system.

Because very different measurement schemes are involved in the above definitions of work, the nature of the relationship  between these is not immediately clear. An analogous question related to heat has been raised by De Roeck et al.~\cite{DeRoeck2007,Derezinski2008,Esposito2009} previously. They proved that, if the open system can be described by the Markovian master equation, these two types of heat are indeed equivalent. Garrahan and Lesanovsky~\cite{Garrahan2010} further emphasized this equivalence from the viewpoint of the evolution equation of the characteristic function (CF) of the heat~\footnote{ Ref.~\cite{Garrahan2010} in fact used the generating function rather than the CF. The former can be viewed as the latter that is evaluated on the imaginary axis. They are no essential differences if the nonequilibrium process is finished within a finite time interval. This notion will be used again in the following discussions}. To my knowledge, however, even when using the Markovian setup, the equivalence of the two types of work has not yet been explicitly discussed, although intuitively this should be true. A possible reason for this is that, under the action of a time-varying force, there are no generally valid Markovian master equations except that the driving force is weak enough~\cite{Breuer2002,Rivas2010}. Very recently, for this specific type of master equation, I developed a CF method~\cite{Liu2014} for calculating the work $W_{QJT}$. It is of note that shortly after the publication of my paper, Silaev et al.~\cite{Silaev2014} proposed another CF method for use in calculating the work $W_{TEM}$ for the same quantum system. Hence, now  it is important to check the equivalence of these two types of work. In this Article, I explicitly show the result.   }\\

{\noindent \it Overview of the two CF methods.} \label{section2}
Let us suppose that the Hamiltonian of a bare system is $H_0$. {This weakly interacts with the surrounding heat reservoir,  with Hamiltonian $H_r$, and the inverse temperature $\beta$. The interaction term is assumed to be $V$$=$$S\otimes$$R$, where $S$ and $R$ represent the operators of the system and reservoir parts, respectively. The form of $V$ is not the most general form possible, although it is adequate for illustrating our results. Initially, the system is in the thermal state,
$\rho_0$$=$$\exp(-\beta H_0)/{\rm Tr}[\exp(-\beta H_0)]$. Then, a weakly driving field is applied from time $t$$=$$0$ to a final time, $t_f$. If one assumes that several approximations are appropriate during the process, which includes the weak-coupling, Born-Markov, and rotation wave approximations (RWA), the evolution equation of the reduced density matrix of the system $\rho(t)$ is (setting $\hbar$$=$$1$)~\cite{Breuer2002}}
\begin{eqnarray}
\label{QMME}
\partial_{t}\rho(t)={\cal L}_{t}\rho(t)=-i[H_0+H_1(t),\rho(t)]+D[\rho(t)],
\end{eqnarray}
where $H_1(t)$ is the interaction term between the system and the field. The $D$-term denotes dissipation due to weak coupling between the system and the reservoir:
\begin{eqnarray}\label{dissipation}
D[A]=\sum_{\omega}\gamma(\omega)\left[S(\omega)A S^\dag(\omega)- \frac{1}{2}
\left\{S^\dag (\omega)S(\omega),A\right\}\right].
\end{eqnarray}
The operator, $S$, can be decomposed into a sum of the eigen-operators of $H_0$, i.e.,
\begin{eqnarray}
S=\sum_\omega S(\omega)=\sum_\omega S^\dag(\omega),
\end{eqnarray}
$[H_0,S(\omega)]$$=$$\textendash\omega S(\omega)$, and $S(-\omega)$$=$$S^\dag(\omega)$. These sums are extended over all energy differences, $\omega$, of the eigenvalues of $H_0$~\cite{Breuer2002}. The rate $\gamma(\omega)$ satisfies the detailed balance
condition:
\begin{eqnarray}
\gamma(-\omega)=\gamma(\omega)\exp(-\beta\omega).
\end{eqnarray}
This condition plays a key role in the validity of the work equalities~\cite{Liu2014,Liu2014b} and in the following discussions. Master equation~(\ref{QMME}) is widely utilized in quantum optics, e.g., in describing resonance fluorescence~\cite{Breuer2002}.

Equation~(\ref{QMME}) can be unraveled into the QJT form for the state vector~\cite{Breuer2002,Carmichael1993,Wiseman2010}. By applying this notion, one may intuitively define the work $W_{QJT}$ along each individual trajectory~\cite{Horowitz2012,Hekking2013,Liu2014}, as described in Fig.(\ref{figure})(b). I have shown that the probability density function (PDF) of the work can be calculated using its CF~\cite{Liu2014}
\begin{eqnarray}
\label{ourcharacterfun} \Phi_{QJT}(\mu)=E[e^{i\mu W_{QJT}}]={\rm
Tr}_0[K_0(0;\mu)\rho_0],
\end{eqnarray}
{where $\mu$ is a real number, and the symbol $E[\cdots]$ denotes an average over all QJTs.}
The operator $K_0(t';u)$ $(0\le t'\le t_f)$ therein satisfies
\begin{eqnarray}
\label{ourevolutionequation}
\partial_{t'}K_0(t';\mu)&=&-{\cal L}^\star_{t'} K_0(t';\mu) -K_0(t';\mu)\hspace{0.1cm}i[H_1(t'),\hspace{0.1cm}e^{i\mu H_0}]e^{-i\mu H_0},\nonumber\\
&=&-i[H_0+H_1(t'),K_0(t';\mu)]\nonumber\\
&&\hspace{1cm}-D^\star[K_0(t';\mu)]-
K_0(t';\mu)\hspace{0.1cm}i[H_1(t'),\hspace{0.1cm}e^{i\mu H_0}]e^{-i\mu H_0},
\end{eqnarray}
where $K_0(t_f;\mu)$$=$$I$ is the identity operator, ${\cal L}^\star_{t'}$ is the dual superoperator of ${\cal L}_{t'}$, and
\begin{eqnarray}
\label{dualdissipation}
D^\star[A]=\sum_{\omega}\gamma(\omega)\left[S^\dag(\omega)A S(\omega)- \frac{1}{2}
\left\{S^\dag (\omega)S(\omega),A\right\}\right].
\end{eqnarray}
Equation~(\ref{ourevolutionequation}) corresponds to the backward time, $t'$. Hence, I refer to this as the backward equation. This is a terminal value problem rather than a common initial value problem. On the other hand, for use with the same master equation, Silaev et al.~\cite{Silaev2014} presented another CF for use with the work, $W_{TEM}$:
\begin{eqnarray}
\label{Silaevcharacterfun} \Phi_{TEM}(\mu)={\rm
Tr}_0[e^{i\mu H_0}\hat{\rho}(t_f;\mu)],
\end{eqnarray}
where a modified reduced density matrix $\hat{\rho}(t_f;\mu)$ that satisfied the following expression~\cite{Esposito2009} was introduced:
\begin{eqnarray}
\label{Silaevevolutionequation}
\partial_t \hat{\rho}(t;\mu)&=&\breve{\cal L}_t(\mu) \hat{\rho}(t;\mu)\nonumber\\
&=&-i[H_0+H_1(t),\hat{\rho}(t;\mu)]\nonumber\\
&&+\sum_{\omega}  \gamma(\omega)\left[
e^{i\mu\omega}S(\omega) \hat{\rho}(t;\mu)S^\dag(\omega)-\frac{1}{2}\left\{S^\dag(\omega)S(\omega),\hat{\rho}(t;\mu)\right\}\right],
\end{eqnarray}
with an initial condition of $\exp[-i\mu H_0]\rho_0$. In order to distinguish this equation from Eq.~(\ref{ourevolutionequation}), I refer to this as the forward equation, since it relates to the forward time, $t$. The forms of Eqs.~(\ref{ourcharacterfun}) and~(\ref{Silaevcharacterfun}), and Eqs.~(\ref{ourevolutionequation})  and~(\ref{Silaevevolutionequation}) appear to be very distinct. \\

{\noindent \it Directly proving the equivalence of the two methods.}
\label{section3}
First, I introduce an alternative operator,
\begin{eqnarray}
\widetilde{K}_0(s;\mu)=\Theta K_0(t';\mu)e^{i\mu H_0}\Theta^\dag,
\end{eqnarray}
for which $\Theta$ is the time-reversal operator and a parameter $s$$=$$t_f-t'$. By substituting this expression into Eq.~(\ref{ourevolutionequation}), I convert this equation into an initial value problem:
\begin{eqnarray}
\label{ourevolutionequation2}
\partial_s \widetilde{K}_0(s;\mu)&=&\widetilde{\breve{{\cal L}}}_s(\mu) \widetilde{K}_0(s;\mu)\nonumber\\
&=&-i[H_0+\widetilde{H}_1(s),\widetilde{K}_0(s;\mu)]\nonumber\\
&& +\sum_{\omega}\gamma(\omega)\left[e^{-i\mu\omega}S^\dag(\omega)\widetilde{K}_0(s;\mu)S(\omega)- \frac{1}{2}
\left\{S^\dag (\omega)S(\omega),\widetilde{K}_0(s;\mu)\right\}\right],
\end{eqnarray}
where $\widetilde{H}_1(s)$$=$$\Theta H_1(t_f-s)\Theta^{\dag}$, and the initial condition $\widetilde{K}_0(0;\mu)$ is equal to $\exp[-i\mu H_0]$.
I have used the detail balance condition and assumed that the eigenoperators, $S(\omega)$, are time-reversible. The formal solution to Eq.~(\ref{ourevolutionequation2}) is given by the following expression:
\begin{eqnarray}
\widetilde{K}_0(s;\mu)=\widetilde {\breve{G}}(s,0;\mu)[\widetilde{K}_0(0;\mu)],
\end{eqnarray}
where $\widetilde {\breve{G}}(s,0;\mu)$$=$$T_- \exp\left[{\int_0^s}\widetilde{\breve{{\cal L}}}_{s'}(\mu)ds'\right]$ is a superpropagator and $T_-$ denotes the chronological time-ordering operator. The superoperator $\widetilde{\breve{{\cal L}}}_s(\mu)$ possesses an important property:
\begin{eqnarray}
\label{equality}
\widetilde{\breve{{\cal L}}}_s(\mu)A= \Theta\breve{{\cal L}}^\star_{t'}(\mu)(\Theta^\dag A \Theta) \Theta^\dag,
\end{eqnarray}
where $\breve{\cal {L}}^\star_t(\mu)$ is the dual superoperator of $\breve{\cal {L}}_t(\mu)$ in Eq.~(\ref{Silaevevolutionequation}):
\begin{eqnarray}
\breve{\cal {L}}^\star_t(\mu)A=i[H_0+H_1(t),A]+
\sum_{\omega}\gamma(\omega)\left[e^{i\mu\omega}S^\dag(\omega)AS(\omega)- \frac{1}{2}
\left\{S^\dag (\omega)S(\omega),A\right\}\right].
\end{eqnarray}
The validity of Eq.~(\ref{equality}) can be easily proved. Using these operators and the superoperator property mentioned above, I rewrite Eq.~(\ref{ourcharacterfun}) using the following expressions:
\begin{eqnarray}
\Phi_{QJT}(\mu)&=&{\rm
Tr}[\Theta^\dag  \widetilde{K}_0(t_f;\mu)\Theta e^{-i\mu H_0}\rho_0]\\
&=&{\rm Tr}[\Theta^\dag \widetilde{\breve{G}}(t_f,0;\mu)[\widetilde{K}_0(0;\mu)]\Theta e^{-i\mu H_0}\rho_0]\\
&=&{\rm Tr}[\breve{G}^\star (0,t_f;\mu)[\Theta^\dag \widetilde{K}_0(0;\mu)\Theta]  e^{-i\mu H_0}\rho_0]\\
&=&{\rm Tr}[e^{i\mu H_0}\breve{ G} (t_f,0;\mu)[e^{-i\mu H_0}\rho_0]],
\label{transformingCF }
\end{eqnarray}
where $\breve{ G}(t_f,0;\mu)$ is the dual superoperator of $\breve{G}^\star (0,t_f;\mu)$$=$$T_+ \exp\left[{\int_0^{t_f}}\breve{{\cal L}}^\star_{\tau}(\mu)d\tau\right]$, and $T_+$ denotes
the antichronological time-ordering operator. It is clear that $\breve{G}(t_f,0;\mu)$ is in fact the superpropagator contained in Eq.~(\ref{Silaevevolutionequation}). {Therefore, these two CF methods are equivalent, i.e., $\Phi_{QJT}(\mu)$$=$$\Phi_{TEM}(\mu)$.} Since the backward and forward time parameters are involved, the current situation is very similar to the case of the Kolmogorov backward and forward equations employed in classical stochastic theory~\cite{Risken1984,Liu2010}.\\
\\

{\noindent \it  Microscopic basis for the backward equation.}
\label{section4}
Equation~(\ref{Silaevevolutionequation}) has a microscopic origin. This expression was obtained by reducing an equation relating to the CF of the work, $W_{TEM}$, defined for the composite system~\cite{Esposito2009,Campisi2011,Silaev2014}, into the degrees of freedom of the system. The above proof implies that Eq.~(\ref{ourevolutionequation}) shall be derived in an analogous manner, even though this expression was obtained solely by employing the notion of QJT~\cite{Liu2014}. After all, the quantum master equation and its manner of unraveling can be thought of as effective theories. {However, if one follows the forward approach, as employed by Silaev et al., it is impossible to directly arrive at the backward equation~(\ref{ourevolutionequation}). Let us recall another CF method for computing the PDF of the work, $W_{TEM}$, of the composite system. First, I explicitly write its whole Hamiltonian, $H(t')$$=$$H_0$$+$$H_1(t')$$+$$H_r$$+$$V$.} There exists the following evolution equation relating to the operator, $K(t';\mu)$~\cite{Liu2014b}:
\begin{eqnarray}
\label{closedevolutionequation}
\partial_{t'}K(t' ;\mu)=-i[H(t'),K(t';\mu)]-
K(t';\mu)i[H(t'), e^{i\mu(H_0+H_r)}]e^{-i\mu(H_0+H_r)}
\end{eqnarray}
with a terminal condition of $K(0;\mu)$$=$$I$. The CF, $\Phi_{TEM}(\mu)$, equals $ {\rm Tr}[K(0;\mu)\rho_0\otimes\rho_r]$, where $\rho_r$$=$$\exp(-\beta H_r)/{\rm Tr}[\exp(-\beta H_r)]$ is the canonical density matrix of the reservoir.
A brief explanation of Eq.(\ref{closedevolutionequation}) is given in Appendix I. This equation appears to be very similar to Eq.~(\ref{ourevolutionequation}). In particular, as $\Phi_{TEM}(\mu)$$=$$ {\rm Tr}_0[{\rm Tr}_r[K(0;\mu)\rho_r]\rho_0]$, one may naturally assert that the latter could correspond to the reduced effective equation of the former, while the term ${\rm Tr}_r[\cdots]$ (the trace over the reservoir) could correspond to the previous $K_0(0;\mu)$. In the following discussion, these two conjectures are verified.

Let us introduce the time evolution operator $U_{0r}(t')$$=$$\exp[ -i(H_0+H_r)t']$ and rewrite Eq.~(\ref{closedevolutionequation}) using the interaction picture,
\begin{eqnarray}
\label{eqinteractionPic}
&&\partial_{t'}K_I(t';\mu)+i[H_1^I(t'),K_I(t';\mu)]+iK_I(t';\mu)e^{i\mu H_0}[e^{-i\mu H_0},H_1^{I}(t')]\nonumber\\
&=&-i[V_I(t')K_I(t';\mu)-K_I(t';\mu)V_{I{\mu}}(t')],
\end{eqnarray}
where the operator subscripts {\it I} denote that these are the interaction picture operators, and
\begin{eqnarray}
V_{I{\mu}}(t')=U^\dag_{0r}(t')e^{i\mu(H_0+H_r)}Ve^{-i\mu(H_0+H_r)}U_{0r}(t').
\end{eqnarray}
Note that I have moved all terms that do not involve the interaction term, $V$, to the left-hand side (LHS) of the equation. This is in preparation for the perturbation calculation below. Eq.~(\ref{eqinteractionPic}) has an integral form:
\begin{eqnarray}
\label{eqinteractionPicintegral}
K_I(t';\mu)=I&&+i\int_{t'}^{t_f}d\tau[V_I(\tau)K_I(\tau;\mu)-K_I(\tau;\mu)V_{I{\mu}}(\tau)]\nonumber\\
&&+i\int_{t'}^{t_f}d\tau [H_1^I(\tau),K_I(\tau;\mu)]+i\int_{t'}^{t_f}d\tau K_I(\tau;\mu)e^{i\mu H_0}[e^{-i\mu H_0},H_1^I(\tau)].
\end{eqnarray}
I substitute Eq.~(\ref{eqinteractionPicintegral}) into the right-hand side (RHS) of Eq.~(\ref{eqinteractionPic}) and obtain
\begin{eqnarray}
\label{eqinteractionPicexpand}
&&\partial_{t'}K_I(t';\mu)+i[H_1^I(t'),K_I(t';\mu)]+iK_I(t';\mu)e^{i\mu H_0}[e^{-i\mu H_0},H_1^I(t')]\nonumber\\
&&=\int_{t'}^{t_f}d\tau V_I(t')[V_I(\tau)K_I(\tau;\mu)-K_I(\tau;\mu)V_{I{\mu}}(\tau)] \nonumber\\
&&\hspace{0.5cm}-\int_{t'}^{t_f}d\tau [V_I(\tau)K_I(\tau;\mu)-K_I(\tau;\mu)V_{I{\mu}}(\tau)]V_{I\mu}(t') -i[V_I(t')-V_{I{\mu}}(t')]\nonumber\\
&&\hspace{0.5cm}+\int_{t'}^{t_f}d\tau V_I(t')[H_1^I(\tau),K_I(\tau;\mu)]-\int_{t'}^{t_f}d\tau [H_1^I(\tau),K_I(\tau;\mu)]V_{I\mu}(t')\nonumber\\
&&\hspace{0.5cm}+\int_{t'}^{t_f}d\tau V_I(t')K_I(\tau;\mu)e^{i\mu H_0}[e^{-i\mu H_0},H_1^I(\tau)]\nonumber\\
&&\hspace{0.5cm}-\int_{t'}^{t_f}d\tau K_I(\tau;\mu)e^{i\mu H_0}[e^{-i\mu H_0},H_1^I(\tau)]V_{I\mu}(t').
\end{eqnarray}
Multiplying both sides by $\rho_r$ and taking a trace over the reservoir, I transform the LHS of the above equation into
\begin{eqnarray}
\label{eqinteractionPicexpandL}
LHS=\partial_{t'}K_{0I}(t';\mu)
+i[H_1^{I}(t'),K_{0I}(t';\mu)]+iK_{0I}(t';\mu)e^{i\mu H_0}[e^{-i\mu H_0},H_1^{I}(t')],
\end{eqnarray}
where $K_{0I}(t';\mu)$$=$${\rm Tr}_r[K(t';\mu)\rho_r]$. Note that $H_1^I(t')$ is now the interaction picture operator for the system due to the fact that $[H_1(t'),H_r]$$=$$0$. So far, all these derivations are exact. In order to attend to the complicated RHS of Eq.~(\ref{eqinteractionPicexpand}), however, I resort to certain approximations. Following the standard concepts pertaining to the dynamics of open systems~\cite{Breuer2002}, I make an important assumption, that $K_I(\tau;\mu)$$\approx$$K_{0I}(\tau;\mu)$$\otimes $$I_r$ for all $K_I$-terms on the RHS. This is justified if the field term, $H_1(t)$, and the coupling term, $V$, are so very weak that the reservoir is almost not affected by these interactions. By imposing the conventional condition ${\rm Tr}[R_I(t')\rho_r]$$=$$0$ and performing a Markov approximation, I immediately obtain the following expression:
\begin{eqnarray}
\label{eqinteractionPicexpandR}
RHS&=&\int_{t'}^{t_f}d\tau S_{I}(t')S_{I}(\tau)K_{0I}(t';\mu)\langle  R_{I}(t')R_{I}(\tau)\rangle_r\nonumber\\
&&-\int_{t'}^{t_f}d\tau S_I(t')K_{0I}(t';\mu)S_{I\mu}(\tau)\langle R_I(t')R_{I\mu}(\tau)\rangle_r \nonumber\\
&&-\int_{t'}^{t_f}d\tau S_I(\tau)K_{0I}(t';\mu)S_{I\mu}(t')\langle R_I(\tau)R_{I\mu}(t')\rangle_r\nonumber\\ &&+\int_{t'}^{t_f}d\tau K_{0I}(t';\mu)S_{I{\mu}}(\tau)S_{I\mu}(t')\langle R_{I{\mu}}(\tau)R_{I\mu}(t')\rangle_r, \end{eqnarray}
where $\langle\cdots\rangle_r$$=$${\rm Tr}_r[\cdots\rho_r]$ are the reservoir correlation functions. If I further assume the RWA to be valid, the integrals in the above equation may be eliminated, and I arrive at a final form:
\begin{eqnarray}
\label{eqinteractionPicexpandR2}
RHS=-i[H_{LS},K_{0I}(t';\mu)]- D^\star[K_{0I}(t';\mu)],
\end{eqnarray}
where $H_{LS}$ represents the Lamb shift~\cite{Breuer2002}. Considering that this is a standard procedure, I only highlight the key steps in Appendix II. If I transform Eqs.~(\ref{eqinteractionPicexpandL}) and~(\ref{eqinteractionPicexpandR2}) back into Schr\"{o}dinger's theoretical framework and neglect the smaller Lamb shift, I reproduce Eq.~(\ref{ourevolutionequation}). \\
\\

{{\noindent \it The backward equation of the heat.} As discussed earlier, the PDFs of the heat $Q_{TEM}$ and $Q_{QJT}$ are the same in the master equation~(\ref{QMME}). Hence, defining one CF for the heat is adequate, e.g., $\Xi(\mu)$. Previous results~\cite{Derezinski2008,DeRoeck2007,Esposito2009,Garrahan2010,Silaev2014} have shown that this function can be evaluated by
\begin{eqnarray}
\label{characteristicfunheatforward}
\Xi(\mu)=E[e^{i\mu Q_{QJT}}]={\rm Tr}_0[\hat\rho(t_f;\mu)],
\end{eqnarray}
where $\hat\rho$ in the latter equation satisfies the same Eq.~(\ref{Silaevevolutionequation}) but with a different initial condition of $\rho_0$. In comparison with the case of the work, one may expect that there exists an operator $F_0(t';\mu)$ $(0\le t'\le t_f)$ that leads to the relation
\begin{eqnarray}
\label{characteristicfunheatbackward}
\Xi(\mu)={\rm Tr}[F_0(0;\mu)\rho_0].
\end{eqnarray}
It is not difficult to find the backward equation of the operator if one follows either of the two methods related to the work. Here I only present the final result:
\begin{eqnarray}
\label{heatevolutionequation}
\partial_{t'}F_0(t';\mu)&=&-\breve{\cal L}^\star_{t'}(\mu) F_0(t';\mu)
\end{eqnarray}
with a terminal condition $F_0(t_f;\mu)=I$. The current discussion also highlights the fact that, if the initial density matrix of the system is a completely random ensemble, $\rho_C$, such as in the $N$-level system, $\rho_C$$=$$I/N$,
the PDF of the heat must obey an exact equality,
\begin{eqnarray}
\label{heatequality}
E_C[e^{-\beta Q_{QJT}}]=1,
\end{eqnarray}
where I used the subscript $C$ to indicate the fact that all the QJTs star from this specific initial condition. The proof of this is reserved for Appendix III. Note that there exists a classical version of the equality in the classical stochastic process~\cite{Liu2010}.

I would like to assert that, since the use of the backward and forward equations related to the work (heat) always leads to the same CFs, from the viewpoint of computing, there are no specific reasons to favor one more than the other. This is only a question of personal taste or convention. On the other hand, when one carries out formal derivations, in some cases the forward method is more convenient than the backward method, or vice versa. For instance, one can easily obtain a concise expression related to the work equality if the backward method is used~\cite{Liu2014a,Liu2014,Liu2014b}.}
\\

{\noindent \it Conclusion.}
\label{section5}
{In this paper, I have proved the equivalence of the two CF methods used for calculating work in weakly driven open quantum systems. Hence, in the present case, the PDF of the work defined using QJT is the same as the PDF of the work defined for the combined system and reservoir using the TEM scheme. This finding has two implications. First, conceptually, the definition of the work and heat using the QJT is intuitive if one wants to understand these thermodynamic quantities from the viewpoint of the system. It is also not in conflict with the definitions that employ the TEM scheme on the system and reservoir. Second, since the recording of the QJTs has been realized in experiments~\cite{Nagourney1986,Murch2013}, it shall be realistic in practice to verify the work or heat equalities based on this notion. In addition, my current results could be extended to some specific quantum master equations. For instance, an immediate example is the adiabatically driven quantum master equations~\cite{Liu2014a}. Although concrete analyses are still to be conducted, the high similarity between this type of master equation and the current version strongly suggests this conjecture. Finally, I want to emphasize that conducting the TEM scheme on an arbitrary composite system is always possible, as this does not rely on the coupling strength or the time-evolution scales between the system and reservoir. However, whether or not QJTs exist for general cases, e.g., strongly coupling and nonMarkovian conditions, is still an unresolved issue. Hence, under these conditions, defining work or heat from the viewpoint of the system would become very challenging. I hope that new progresses in this direction can be presented in the near future. }\\
\\

{\noindent \it Acknowledgment.} I thank Xin Wang for his participation in helpful discussions. I would also like to thank Editage (http://www.editage.com) for English language editing. The work was supported by the National Science Foundations of China under Grant No. 11174025 and No. 11575016.

\section*{Appendix I: An explanation of Eq.~(\ref{closedevolutionequation})}
I discuss here the time-dependent Hamiltonian of a closed quantum system, $H(t)$$=$$H_s$$+$$H_1(t)$. This Hamiltonian, $H_s$, is assumed to have discrete eigenstates and eigenvalues: $H_s|n\rangle$$=$$\epsilon_n|n\rangle$. According to the TEM scheme~\cite{Campisi2011}, one may define the exclusive work as $W_{TEM}$$=$$\epsilon_n$$-$$\epsilon_m$, where$\epsilon_i$, $i$$=$$m,n$ denotes the energy eigenvalues of $H_s$ that are measured at the beginning and the end of the nonequilibrium process. The CF of the PDF of the work is $\Phi(\mu)$$=$$ {\rm Tr}[K_{t_f}(\mu)\rho_s]$, where
\begin{eqnarray}
  K_t(\mu)=U^\dag(t)e^{i\mu H_s} U(t) e^{-i\mu H_s},
\end{eqnarray}
$U(t)$ is the time-evolution operator of $H(t)$, and  $\rho_s$ is the thermal density matrix of the bare system, $H_s$. One way to calculate $\Phi(\mu)$ is to directly solve $U(t)$. An alternative method involves determining an evolution equation about $K_t(\mu)$. However, there is no such closed equation about $K_t(\mu)$ with respect to the time, $t$. This problem may be circumvented by introducing
\begin{eqnarray}
K(t';\mu)=U(t')U^\dag(t_f)e^{i\mu H_s} U(t_f)U^\dag(t')e^{-i\mu H_s}.
\end{eqnarray}
It is obvious here that $K(0;\mu)$$=$$K_{t_f}(\mu)$ and $K(t_f;\mu)$$=$$I$. Interestingly, this new operator satisfies the following closed evolution equation about the backward time, $t'$,
\begin{eqnarray}
\partial_{t'}K(t' ;\mu)=-i[H(t'),K(t';\mu)]-
K(t';\mu)i[H(t'), e^{i\mu H_s}]e^{-i\mu H_s}.
\end{eqnarray}
If the bare system is composed of the aforementioned combined system and reservoir, and the interaction term, $V$, is assumed to be negligible, I arrive at Eq.~(\ref{closedevolutionequation}).

\section*{Appendix II: Several key formulas used in deriving Eq.~(\ref{eqinteractionPicexpandR2})}
The decomposition of the operator $S$ implies that
\begin{eqnarray}
S_{I\mu}(t)=\sum_{\omega} S(\omega )e^{-i\omega (t+\mu)}=\sum_\omega S^\dag(\omega) e^{i\omega(t+\mu)}.
\end{eqnarray}
Substituting these into Eq.~(\ref{eqinteractionPicexpandR}) and performing RWA, I obtain
\begin{eqnarray}
\label{details}
RHS&=&\sum_\omega S(\omega)S^\dag(\omega) K_{0I}(t';\mu)\int_{0}^{t_f-t'}ds e^{i\omega s}\langle  R_{I}(0)R_{I}(-s)\rangle_r\nonumber\\
&&-\sum_\omega S(\omega)K_{0I}(t';\mu)S^\dag(\omega) e^{i\mu\omega}\int_{0}^{t_f-t'}ds  e^{i\omega s}\langle R_I(0)R_{I\mu}(s)\rangle_r \nonumber\\
&&-\sum_\omega S(\omega) K_{0I}(t';\mu)S^\dag(\omega) e^{i\mu\omega}\int_{0}^{t_f-t'}ds e^{-i\omega s}\langle R_I(0)R_{I\mu}(-s)\rangle\nonumber\\
&&+\sum_\omega K_{0I}(t';\mu)S(\omega) S^\dag(\omega) \int_{0}^{t_f-t'}ds e^{-i\omega s}\langle R_{I{\mu}}(0)R_{I\mu}(-s)\rangle_r.
\end{eqnarray}
If the correlation functions decay very quickly, these integrals can be approximated using one-sided Fourier transforms by replacing the upper limit $t_f$$-$$t'$ with infinity. Furthermore, it is useful to rewrite these one-sided Fourier transforms using the positive double-sided Fourier transforms $\gamma(\omega)=\int_{-\infty}^{+\infty}e^{i\omega s}\langle R_I(0)R_I(-s) \rangle_r$~\cite{Breuer2002}:
\begin{eqnarray}
\int_{0}^{\infty}ds e^{i\omega s}\langle  R_{I}(0)R_{I}(-s)\rangle_r&=&\frac{1}{2}\gamma(\omega)+\frac{i}{2\pi}{\cal P}\int_{-\infty}^{+\infty}\frac{\gamma (\Omega)}{\omega-\Omega}d\Omega, \\
\int_{0}^{\infty}ds e^{i\omega s}\langle  R_{I}(0)R_{I\mu}(-s)\rangle_r&=&\frac{1}{2}\gamma(\omega)e^{i\omega\mu}+\frac{i}{2\pi}{\cal P}\int_{-\infty}^{+\infty}\frac{\gamma (\Omega)}{\omega-\Omega}e^{i\Omega\mu}d\Omega,
\end{eqnarray}
where $\cal P$ denotes the Cauchy principal value
of the integral. After a simple rearrangement, I obtain Eq.~(\ref{eqinteractionPicexpandR2}), where the Lamb shift term
\begin{eqnarray}
H_{LS}=\sum_\omega S^\dag(\omega) S(\omega)(1/2\pi)P\int_{-\infty}^{+\infty}{\gamma(\Omega)}/(\omega-\Omega)d\Omega.
\end{eqnarray}

{
\section*{Appendix III: Proof of Eq.~(\ref{heatequality})}
Two key observations are used here. First, Eqs.~(\ref{characteristicfunheatforward}) and~(\ref{characteristicfunheatbackward}) are in fact valid for any initial density matrix that is diagonalized in terms of the energy basis~\cite{Esposito2009,Liu2014a}, e.g.,
\begin{eqnarray}
\label{equalityrandommatrix}
E_C[e^{i\mu Q_{QJT}}]={\rm Tr}[F_0(0;\mu)\rho_C]={\rm Tr}_0[\hat\rho(t_f;\mu)].
\end{eqnarray}
The modified reduced density matrix $\hat\rho(t;\mu)$ still satisfies Eq.~(\ref{Silaevevolutionequation}) but with the initial density matrix of $\rho_C$. Second, by applying the detailed balance condition about the rates and $S(-\omega)$$=$$S^\dag(\omega)$, one may find
\begin{eqnarray}
\label{propertyL}
\breve{\cal L}_{t'}(i\beta)(\rho_C)=0.
\end{eqnarray}
With these two facts, I can easily prove the equality by using the backward equation~(\ref{heatevolutionequation}) or the forward equation~(\ref{Silaevevolutionequation}). For the former method, I need to evaluate the following expression:
\begin{eqnarray}
\label{proveheatequality}
\partial_{t'}{\rm Tr}[F_0(t';\mu)\rho_C]&=&-{\rm Tr}[\breve{\cal L}^\star_{t'}(\mu)( F_0(t';\mu))\rho_C]\nonumber\\
&=&-{\rm Tr}[F_0(t';\mu)\breve{\cal L}_{t'}(\mu)(\rho_C)]
\end{eqnarray}
Clearly, by substituting $\mu$$=$$i\beta$ into the above equation, performing an integral on both sides from time $0$ to $t_f$, and applying the terminal condition $F_0(t_f;\mu)$$=$$I$, I immediately obtain the identity,
\begin{eqnarray}
{\rm Tr}[F_0(0;i\beta)\rho_C]=1.
\end{eqnarray}
According to Eq.~(\ref{equalityrandommatrix}), I may then prove Eq.~(\ref{heatequality}). However, it is also easy to arrive at this equality if one applies the forward equation~(\ref{Silaevevolutionequation}). I first re-express Eq.~(\ref{equalityrandommatrix}) as
\begin{eqnarray}
{\rm Tr}_0[\hat\rho(t_f;\mu)]={\rm Tr}[\breve{ G} (t_f,0;\mu)[\rho_C]],
\label{transformingCF }
\end{eqnarray}
where the superpropagator $\breve{G}(t_f,0;\mu)$ equals $T_- \exp\left[{\int_0^{t_f}}\breve{{\cal L}}_{\tau}(\mu)d\tau\right]$.  Choosing $\mu$$=$$i\beta$ and examining Eqs.~(\ref{equalityrandommatrix}) and~(\ref{propertyL}) again, I re-obtain the equality~(\ref{heatequality}). The reader is reminded that these discussions are also able to verify the validity of the work equalities~\cite{Liu2014,Liu2014a}.}
\bibliography{RFsubmission}.

\end{document}